\documentclass[pra,superscriptaddress,showpacs,amssymb,letterpaper,twocolumn]{revtex4}

\usepackage{amsmath}
\usepackage{amsfonts}
\usepackage{graphicx}

\def\>{\rangle}
\def\<{\langle}
\def\be{\begin{equation}}
\def\ee{\end{equation}}
\def\qed{\leavevmode\unskip\penalty9999 \hbox{}\nobreak\hfill
     \quad$\blacksquare$
     \par\vskip3pt}

\begin{document}
\title{Unambiguous discrimination of mixed states}

\author{Terry Rudolph}
%\email{rudolpht@bell-labs.com}
\affiliation{Bell Labs, 600 Mountain Ave., Murray Hill, NJ 07974,
U.S.A.}

\author{Robert W. Spekkens}
%\email{spekkens@perimeterinstitute.ca}
\affiliation{Perimeter Institute for Theoretical Physics, 35 King
St. North, Waterloo, Ontario N2J 2W9, Canada}
\affiliation{Department of Physics, University of Toronto, 60
St.George Street, Toronto, Ontario M5S 1A7, Canada}
\author{Peter S. Turner}
%\email{psturner@physics.utoronto.ca}
\affiliation{Department of Physics, University of Toronto, 60
St.George Street, Toronto, Ontario M5S 1A7, Canada}

\date{\today}

\begin{abstract}
We present the conditions under which probabilistic error-free
discrimination of mixed states is possible, and provide upper and
lower bounds on the maximum probability of success for the case of
two mixed states. We solve certain special cases exactly, and
demonstrate how the problems of state filtering and state
comparison can be recast as problems of mixed state unambiguous
discrimination.

\end{abstract}


\maketitle

\section{Introduction}

A characteristic feature of quantum mechanics is that if a system
is prepared in one of a set of non-orthogonal pure states, then
there is no measurement that can yield an error-free determination
of which state was prepared. Nonetheless, it \emph{is} possible to
achieve a \emph{probabilistic} error-free discrimination, that is,
one which sometimes fails but when successful never gives an
erroneous result. This sort of discrimination procedure is
generally referred to as \emph{unambiguous discrimination} (UD).
The UD of two \emph{pure} states prepared with equal prior
probabilities was considered by Ivanovic and Dieks \cite{Ivanovic}
and the optimal procedure was given by Peres \cite{Peres}. This
was generalized to the case of unequal prior probabilities by
Jaeger and Shimony \cite{JaegerShimony}. The problem of three pure
states was analysed in \cite{PeresTerno}, while multiple pure
states were considered in \cite{Chefles,Eldar0}.

It is a common misconception that the unambiguous discrimination
of \emph{mixed} states is impossible \cite{FJ,Eldar}. Indeed, in
\cite{FJ} it is explicitly stated that ``one cannot unambiguously
discriminate mixed states." That such a claim cannot be correct
for an arbitrary mixed-state ensemble is proven, however, by the
following counterexample: any set of {\em orthogonal} mixed states
can always be discriminated with zero probability of error.

What is less obvious is that there exist sets of {\em
non-orthogonal } mixed states for which UD is possible. The
critical feature of such sets is that their elements do not have
identical supports \footnote{The \emph{support} of a mixed state
is the space spanned by its eigenvectors; the \emph{kernel} is the
space orthogonal to its support.}. In fact, all that is required
for there to be a non-zero probability of error-free
discrimination is that one of the density operators have a
non-zero overlap with the intersection of the kernels of the
others.

In this paper, we consider the problem of determining the
\emph{optimal} UD procedure for an arbitrary pair of mixed states.
We derive strong upper and lower bounds on the probability of a
conclusive result, and we provide an exact solution in the special
case where both states have kernels of dimension 1.

The ability to unambiguously discriminate a pair of mixed states
has many applications. Indeed, several recently studied problems
can be recast as special cases of UD of mixed states. We shall
focus on two tasks of this sort: \emph{state comparison}
\cite{BCJ} (determining whether two systems are described by the
same or different pure states) and \emph{state filtering}
\cite{SBH,bergou} (discriminating one pure state from a set of
pure states). We find that for the state comparison problem, our
upper and lower bounds coincide, and thus automatically yield the
optimal solution. For the state filtering problem, we find that
our lower bound is equal to the optimal solution found in
\cite{SBH,bergou}. Given that our lower bound has a simple
geometric interpretation, it serves to clarify the nature of the
optimal state filtering procedure. In particular, it shows that
the eigenbases of the optimal POVM elements depend only on the
subspace spanned by the states against which one is trying to
filter, and not on the specific states themselves. In addition,
our approach generalizes in a straightforward way to more
complicated estimation tasks, such as state filtering and state
comparison when the unknown states are themselves mixed.

It should be noted that our results apply not only to mixtures
that arise from ignorance about which of several pure state
descriptions applies, but also to those arising as the reduced
density operator of an entangled state. As such, our results can
be applied to the task of achieving an UD of two entangled states
of a composite system given access to only one of the subsystems.
It is also worth noting that our lower bound is obtained by making
implicit use of the CS decomposition, which constitutes a powerful
tool in both modern linear algebra and classical signal analysis
\cite{CSdecomp}. To our knowledge, this is the first application
of the CS decomposition in \emph{quantum} signal analysis.

\section{\bf General Formulation}

We consider the task of discriminating unambiguously between two
mixed states $\rho _{0}$ and $\rho _{1}$ with prior probabilities
$p_{0}$ and $p_{1}.$ The measurement procedure can have up to
three outcomes, associated with identifying the state as
$\rho_{0}$, identifying the state as $\rho_{1}$, and failing to
identify the state conclusively. The most general three-outcome
measurement is represented in quantum mechanics by a three-element
positive-operator valued measure (POVM), which we denote by
$\{E_{0},E_{1},E_{?}\}$. Because the identification must never be
in error, we require that
\begin{equation}
{\rm Tr}(\rho _{0}E_{1})={\rm Tr}(\rho _{1}E_{0})=0. \label{cons1}
\end{equation}
The probability $P$ of successful UD is:
\begin{equation}
P=p_{0}{\rm Tr}(\rho _{0}E_{0})+p_{1}{\rm Tr}(\rho _{1}E_{1}).
\label{optimize}
\end{equation}

We shall denote the support of $\rho _{b}$ by ${\rm
supp}(\rho_{b})$ and its kernel by ${\cal K}_{b}$ ($b$=0,1). It is
clear that any intersection of ${\cal K}_{0}$ and ${\cal K}_{1}$
is not useful for the purposes of discriminating $\rho _{0}$ and
$\rho _{1}$ since neither state has any overlap with this
subspace. We shall therefore assume, in what follows, that the
Hilbert space is equal to the span of the supports of $\rho _{0}$
and $\rho _{1}$. A necessary and sufficient condition for
satisfying (\ref{cons1}) is that the POVM element $E_{0}$($E_{1}$)
have support only in the subspace ${\cal K}_{1}$(${\cal K}_{0}$).
It follows that for there to be a non-zero probability of success,
at least one of ${\cal K}_{0}$ and ${\cal K} _{1}$ must be
non-zero.  This occurs if and only if ${\rm supp}(\rho _{0})\ne
{\rm supp}(\rho _{1})$.

We seek to maximize (\ref{optimize}) subject to (\ref{cons1}) and
the constraint that $E_{0}$, $E_{1}$, and $E_{?}$ be positive and
sum to the identity.  It suffices to vary over positive $E_{0}$
and $E_{1}$ which satisfy
\begin{equation}
I-E_{0}-E_{1}\geq 0.  \label{cons2}
\end{equation}
This optimization problem is an instance of a semi-definite
programming problem for which there exist efficient numerical
algorithms.  (See \cite{Eldar0} for applications of semi-definite
programming to UD of pure states, and \cite{Eldar,FJ} for
applications to the problem of optimizing the discrimination
probability between mixed states for a fixed error rate, which
interpolates between maximum likelihood estimation and UD).

\section{Solutions in certain special cases}

\subsection{Mixed states with orthogonal kernels}

If ${\cal K}_0 \perp {\cal K}_1$, the optimal POVM which satisfies
(\ref{cons2}) is clearly $E_0=K_{1}$, $E_1=K_{0}$, where $K_b$ is
the projector onto ${\cal K}_b$. This yields a probability of
success $P^{\max}=p_0\mathrm{Tr}(\rho_0 K_1)+p_1\mathrm{Tr}(\rho_1
K_0)$. Note that this solution also applies when one of the
kernels is the null space. Note also that commuting mixed states
necessarily have orthogonal kernels.  It follows that this result
specifies the maximum probability of UD for overlapping classical
probability distributions.

\subsection{Mixed states with 1-dimensional kernels}\label{rank1}

 We now turn to the special case wherein the kernels are both
1-dimensional - that is, the states $\rho _{0}$ and $\rho _{1}$
have rank $n-1,$ and the span of their supports is an $n$
dimensional space. Denoting by $ |
k_{b} \rangle  \langle k_{b} | $ the projector onto ${\cal K}%
_{b}$, the POVM must be of the form $\{E_{0}=\alpha | k_{1}
\rangle \langle k_{1} | ,E_{1}=\beta  | k_{0} \rangle \langle
k_{0} | ,E_{?}=I-\alpha  | k_{1} \rangle \langle k_{1} | -\beta  |
k_{0} \rangle  \langle k_{0} | \}$. Our task therefore becomes to
compute
\begin{equation}
P^{\max }=\max_{\alpha ,\beta }\Big(\alpha p_{0} \langle k_{1} |
\rho _{0} | k_{1} \rangle +\beta p_{1} \langle k_{0} | \rho _{1} |
k_{0} \rangle \Big), \label{rank1opt}
\end{equation}
where the maximization is subject to the constraint
\begin{equation}
I-\alpha  | k_{1} \rangle  \langle k_{1} | -\beta  | k_{0} \rangle
\langle k_{0} | \geq 0. \label{rank1cons2}
\end{equation}
For convenience, we define $\theta $ to be the angle in Hilbert
space between the 1-dimensional kernels,
\begin{equation}
\cos ^{2}\theta \equiv \left| \left\langle
k_{0}|k_{1}\right\rangle \right| ^{2}.  \label{theta}
\end{equation}
Taking the minimum eigenvalue of the left hand side of
(\ref{rank1cons2}), we can re-express this inequality as
\begin{equation}
\frac{1}{2}(\alpha +\beta +\sqrt{(\alpha -\beta )^{2}+4\alpha
\beta \cos ^{2}\theta })\leq 1.  \label{rank1cons3}
\end{equation}
Solving for $P^{\max }$ under this constraint we find that
$\alpha^{\mathrm{opt}}=(1-\sqrt{A_1/A_0}\cos\theta)/\sin^2\theta$,
$\beta^{\mathrm{opt}}=(1-\sqrt{A_0/A_1}\cos\theta)/\sin^2\theta$,
and
\begin{equation}\label{soln1d}
P^{\max }= \Big{\lbrace} \begin{array}{ll} \frac{A_{0}+A_{1}-2\cos
\theta \sqrt{A_{0}A_{1}}}{\sin ^{2}\theta
} & \text{ if }\cos \theta <\sqrt{\frac{A_{\min }}{A_{\max }}} \\
A_{\max } & \text{otherwise.}
\end{array}
\end{equation}
where $A_{0} =p_{0} \langle k_{1} | \rho _{0} | k_{1} \rangle ,
A_{1} =p_{1} \langle k_{0} | \rho _{1} | k_{0} \rangle ,$ and
where $A_{\min}={\min}\{A_0,A_1\},A_{\max}={\max}\{A_0,A_1\}$.
This solution, considered as a function of $| k_{0} \rangle $ and
$ | k_{1} \rangle, $ will be denoted $P_{\text{1D}}^{\max }(|k_{0}
\rangle , | k_{1} \rangle ).$

The problem of unambiguously discriminating two non-orthogonal
pure states $|\psi_0\>$, $|\psi_1\>$ is a special case of the
1-dimensional kernel problem. Although $\cos \theta $ is defined
to be the overlap of the 1-dimensional kernels, clearly
$|\<\psi_0|\psi_1\>|=\cos\theta$, and (\ref{soln1d}) becomes
\begin{equation}\label{puresoln}
P^{\max} = \Big{\lbrace}
\begin{array}{ll}
1-2\sqrt{p_{0}p_{1}} |  \langle \psi _{0}|\psi _{1} \rangle
 | & \text{if } |  \langle \psi _{0}|\psi _{1} \rangle
 | <\sqrt{\frac{p_{\min }}{p_{\max }}}  \\
p_{\max }(1- |  \langle \psi _{0}|\psi _{1} \rangle  |^2) &
\text{otherwise,}
\end{array}
\end{equation}
in agreement with \cite{JaegerShimony}. In the case of equal prior
probabilities, we have $P^{\max }=1-|\<\psi_0|\psi_1\>|$, as
expected from \cite{Peres}.

\section{Lower bound}\label{lowerbound}

We consider a strategy that achieves UD of an arbitrary pair of
mixed states and which is strongly dependent
on the geometrical relationship between the two subspaces ${\cal K}_{0}$ and ${\cal K}_{1}:$\\
\textbf{Theorem} \textit{Consider two arbitrary mixed states
$\rho_0$ and $\rho_1$. Denote the dimensionality of their kernels,
${\cal K}_0$ and ${\cal K}_1$ by $r_0$ and $r_1$, and assume that
$r_0\ge r_1$. There exist orthonormal bases
$\{|k_b^j\>\}_{j=1}^{r_b}$ for ${\cal K}_b (b=0,1)$ such that for
$1\le j\le r_0$, $1\le i\le r_1$,
\be\<k_0^j|k_1^i\>=\delta_{ij}\cos(\theta_j),\ee where the
$\theta_j$ are the canonical angles between ${\cal K}_{0}$ and
${\cal K}_{1}$ \cite{Bhatia}. In this case, the expression
\begin{equation}\label{PL}
P_L=\sum_{j=1}^{r_1} P_{\text{1D}}^{\max }(|k_{0}^j
\rangle,|k_{1}^j \rangle)+\sum_{j=r_1+1}^{r_0}
\<k_0^j|\rho_1|k_0^j\>
\end{equation}
forms a lower bound on the maximum probability of discriminating
unambiguously  between $\rho_0$ and $\rho_1$. }

\textbf{Proof} The proof is constructive. Let $X_{b}$ be any
$n\times r_b$ dimensional matrix whose orthonormal columns span
${\cal K}_{b}.$ Define an $r_b\times r_b$ unitary matrix $U_b$,
and an $r_0\times r_1$ matrix $S$ via a singular value
decomposition \cite{Bhatia} \be X_{0}^{\dagger
}X_{1}=U_0SU_1^{\dagger }. \ee The matrix $S$ is of the form
$S=\left[\begin{matrix}C\\O\end{matrix}\right],$ where $C$ is a
diagonal matrix of the form $C=diag(\cos \theta _{1},...,\cos
\theta _{r_1}),$ $\theta_i\in[0,\pi/2]$, while $O$ is an
$(r_0-r_1)\times r_1$ matrix of 0's. Defining $\theta_j=\pi/2$ for
$j>r_1$, and denoting by $ | k_{b}^{i} \rangle $ the $i$'th column
of $X_{b}U_b$, we have constructed bases for ${\cal K}_0$, ${\cal
K}_1$ which satisfy $
 \langle k_{0}^{j}|k_{1}^{i} \rangle =\delta _{ij}\cos \theta
 _{i},
$
as required in the theorem.

The measurement achieving the lower bound is associated with a
POVM wherein $ E_{0} =\sum_{i=1}^{r_1} \alpha _{i} | k_{1}^{i}
\rangle \langle k_{1}^{i} |,\;\; E_{1} =\sum_{i=1}^{r_0} \beta
_{i} | k_{0}^{i} \rangle  \langle k_{0}^{i} |. $ The constraint
(\ref{cons2}) takes the form
\begin{equation}
I-\sum \alpha _{i} | k_{1}^{i} \rangle  \langle k_{1}^{i} | -\sum
\beta _{i} | k_{0}^{i} \rangle
 \langle k_{0}^{i} | \geq 0.  \label{conslower}
\end{equation}
Since the 2 dimensional subspace spanned by $|k_0^j\>$ and
$|k_1^j\>$ for $1\le j\le r_1$ is orthogonal to all other such
subspaces and is orthogonal to $|k_0^j\>$ for $r_1 < j\le r_0$,
the constraint (\ref{conslower}) reduces to constraints of the
form (\ref{rank1cons2}) for $1\le j\le r_1$, and constraints of
the form $I- \beta _{j} | k_{0}^{j} \rangle \langle k_{0}^{j} |
\geq 0$,
 for $r_1 < j\le r_0$. In this
manner, we have reduced the problem to $r_1$ separate
optimizations of the form already considered in section
\ref{rank1}. Solving each of these yields the first term on the
right hand side of (\ref{PL}) in the theorem. The remaining
$r_0-r_1$ optimizations are achieved by taking $\beta _j=1$, which
yields the second term on the right hand side of (\ref{PL}).\qed

In order to understand the geometry of the eigenbases for the POVM
elements in this lower bound, it is helpful to realize that the
canonical angles $\theta_i$ form the unique geometrical invariants
describing the relationship between two subspaces. They can be
defined iteratively: $\theta_1$ is the smallest angle between any
pair of vectors drawn from ${\cal K}_{0}$ and ${\cal K}_{1}$, and
$|k_0^1\>,|k_1^1\>$ are the corresponding pair of vectors.
$\theta_2$ is the smallest such angle after these two vectors are
removed, and so on. In this way, one obtains a simple geometrical
picture of the measurement achieving the lower bound. We note that
we have not found any example of UD wherein this lower bound is
not optimal. Nonetheless, given that the eigenbases for $E_0,E_1$
depend only on the subspaces spanned by $\rho_0,\rho_1$ and not on
the states themselves, there is no reason to expect it to be
optimal in the general case.

%
%
%Given that the eigenbases for $E_0,E_1$ in this lower bound depend
%only on the subspaces spanned by $\rho_0,\rho_1$ and not on the
%states themselves, it seems unlikely that this lower bound is
%optimal. However, as we shall see, we have encountered no examples
%of a problem for which it is not.

\section{Upper bound}

\textbf{Theorem} \textit{An upper bound on the maximum probability
of unambiguously discriminating two mixed states, $\rho_0$ and
$\rho_1$, is
\be
P^{\max }\le  \Big{\lbrace}
\begin{array}{ll}
1-2\sqrt{p_{0}p_{1}}F(\rho _{0},\rho _{1}) & \text{if }F(\rho_{0},\rho_{1})<\sqrt{\frac{p_{\min }}{p_{\max }}} \\
p_{\max }(1-F(\rho _{0},\rho _{1})^2) & \text{otherwise},
\end{array}
\ee
where $F(\rho _{0},\rho _{1})$ $=Tr | \sqrt{\rho _{0}}\sqrt{\rho _{1}}%
 | $ is the fidelity.}

\textbf{Proof} Let $ | \psi _{b} \rangle $ be a purification of
$\rho _{b}$ \cite{Uhlmann}. Clearly, UD of $|\psi_0\>,|\psi_1\>$
with priors $p_0$,$p_1$ can be achieved with a maximal probability
of success that is greater than or equal to the maximum
probability of success in the UD of $\rho _{0}$,$\rho_{1}$ with
the same priors. This follows from the fact that any
discrimination procedure for the latter task serves also as a
discrimination procedure for the former task. Thus,
\[ P^{\max}(p_{0}\rho _{0},p_{1}\rho _{1})\le
\min_{|\psi_0\>,|\psi_1\>} P^{\max}(p_{0} | \psi _{0} \rangle
\langle \psi _{0} | ,p_{1} | \psi _{1} \rangle \langle \psi _{1} |
).
\]
where the minimization is over all purifications of $\rho_b$. We
know, however, that $P^{\max}(p_{0} | \psi _{0} \rangle \langle
\psi _{0} | ,p_{1} | \psi _{1} \rangle \langle \psi _{1} |)$ is
given by Eq. (\ref{puresoln}).
 Minimizing either of the expressions on the right hand side of
(\ref{puresoln}) requires maximizing $| \langle \psi _{0}|\psi
_{1} \rangle|.$ Uhlmann's theorem \cite{Uhlmann} states that the
maximum overlap between purifications of two density operators is
equal to the fidelity between the density operators. Applying this
observation yields the desired upper bound.\qed

In the case of equal prior probabilities, the upper bound has the
simple form $ P^{\max }\le 1-F(\rho _{0},\rho _{1}). $ Numerical
studies of rank 2 mixed states in a 4 dimensional Hilbert space
indicate that even for randomly chosen $\rho_0$ and $\rho_1$, our
upper and lower bounds are generally very close.

\section{Some applications}

\subsection{State comparison}

In \cite{BCJ}, Barnett, Chefles and Jex introduced the following
problem: given two systems, each of which is in one of two
(generally non-orthogonal) states $\{|\psi_1\>,|\psi_2\>\}$, with
what probability can one determine whether the two systems are in
the same state or in different states? Here, we consider the case
when this determination must be made without error if at all.
Assuming equal likelihood for either possibility, we can clearly
interpret the problem as one of achieving UD of the two mixed
states
\begin{eqnarray}
\rho_0 &=&
\tfrac{1}{2}|\psi_1\psi_1\>\<\psi_1\psi_1|+\tfrac{1}{2}|\psi_2\psi_2\>\<\psi_2\psi_2|\nonumber\\
\rho_1 &=&
\tfrac{1}{2}|\psi_1\psi_2\>\<\psi_1\psi_2|+\tfrac{1}{2}|\psi_2\psi_1\>\<\psi_2\psi_1|.
\end{eqnarray}
as was recognized by the authors of \cite{BCJ}.

These are rank two mixed states in a 4-dimensional space; as such
their kernels are also two dimensional. We let $|\bar{\psi_i}\>$
be the state orthogonal to $|\psi_i\>$, and choose phases such
that $\<\bar{\psi_1}|\bar{\psi_2}\>$ is real. It is clear that
${\cal K}_0$ is spanned by $\{|\bar{\psi_1}\bar{\psi_2}\>,
|\bar{\psi_2}\bar{\psi_1}\>\}$ while ${\cal K}_1$ is spanned by
$\{|\bar{\psi_1}\bar{\psi_1}\>, |\bar{\psi_2}\bar{\psi_2}\>\}$.
Using the techniques of section (\ref{lowerbound}), we find that
the canonical basis vectors in ${\cal K}_0$ are
$|k_0^1\>=|\bar{\psi_1}\bar{\psi_2}\>+|\bar{\psi_2}\bar{\psi_1}\>,$
$|k_0^2\>=|\bar{\psi_1}\bar{\psi_2}\>-|\bar{\psi_2}\bar{\psi_1}\>,$
while those of ${\cal K}_1$ are
$|k_1^1\>=|\bar{\psi_1}\bar{\psi_1}\>+|\bar{\psi_2}\bar{\psi_2}\>,$
$|k_1^2\>=|\bar{\psi_1}\bar{\psi_1}\>-|\bar{\psi_2}\bar{\psi_2}\>$
(all suitably normalized).  Since $\<k_0^2|k_1^2\>=0$, these two
states can be included in their respective POVM elements with
weight 1, i.e. $\alpha _2=1=\beta _2$. The remaining pair of
states have overlap $\cos\theta_1\equiv\<k_0^1|k_1^1\>$, and we
can use the solution for 1-dimensional kernels. Performing this
calculation, we find that our lower bound is $1-F(\rho_0,\rho_1)$.
Since this is equal to our upper bound, we automatically know that
this is optimal, and we have the optimal POVM by construction.

\subsection{State filtering}

A problem that has been considered recently by Sun, Bergou and
Hillery \cite{SBH} is that of unambiguously discriminating whether
a state is $\left| \psi _{1}\right\rangle $ or whether it is in
the set $\{\left| \psi _{2}\right\rangle ,\left| \psi
_{3}\right\rangle \}$ when the prior probabilities of the three
states are $\eta _{1},\eta _{2},$ and $\eta _{3}$ respectively.
This task has been called {\em state filtering}. It is
straightforward to see that it is an instance of UD of mixed
states. Specifically, it is the problem of unambiguously
discriminating \[ \rho _{0} =\left| \psi _{1}\right\rangle
\left\langle \psi _{1}\right|, \; \rho _{1} =\tfrac{\eta
_{2}}{\eta _{2}+\eta _{3}}\left| \psi _{2}\right\rangle
\left\langle \psi _{2}\right| +\tfrac{\eta _{3}}{\eta _{2}+\eta
_{3}}\left| \psi _{3}\right\rangle \left\langle \psi _{3}\right| ,
\] with prior probabilities $p_{0} =\eta _{1}, p_{1} =\eta
_{2}+\eta _{3}.$ It is easy to find our lower bound for this
problem. Since ${\cal K}_{1}$ is 1-dimensional, we associate it
with a unique vector $|k_1\>$. The kernel ${\cal K}_{0}$ is
2-dimensional and so there is flexibility in the basis that
diagonalizes $E_1$. Our lower bound dictates that we use the basis
$|k_0^1\>$,$|k_0^2\>$, where $|k_0^1\>$ is the vector in ${\cal
K}_{0}$ that is maximally parallel to $|k_1\>$, and $|k_0^2\>$ is
the vector in ${\cal K}_{0}$ that is orthogonal to both $|k_1\>$
and $|k_0^1\>$.
%for
%which $\<k_1^1|k_0^2\>=0$ and $\<k_1|k_0^1\>=\cos\theta$, where
%\theta is the canonical angle between ${\cal K}_{0}$ and ${\cal
%K}_{1}$.
Our lower bound is
$
P_L= P_{\text{1D}}^{\max }(|k_{0}^1 \rangle,|k_{1} \rangle)+
\<k_0^2|\rho_1|k_0^2\>
$
Defining the canonical angle $\theta$ by
$\cos\theta=\<k_1|k_0^1\>$ and appealing to the geometry of the
problem, we find $A_{0}=p_{0}(1-\cos ^{2}\theta )$ and
$A_{1}=p_{1}F(\rho _{0},\rho _{1})^{2}\tan ^{2}\theta $. Finally,
defining $\tilde{F}\equiv F(\rho _{0},\rho
_{1})/\sqrt{p_{0}/p_{1}}$, the lower bound takes the following
form:
\[
P^{\max}\ge\Bigg\lbrace
\begin{array}{cr}
1-p_{0}\cos ^{2}\theta -\tfrac{p_{1}}{\cos^2\theta}F(\rho
_{0},\rho _{1})^{2} & \tilde{F}\le\cos ^{2}\theta\\
1-2\sqrt{p_{0}p_{1}}F(\rho _{0},\rho _{1})) & \cos ^{2}\theta \le
\tilde{F}\le 1\\
p_{1}(1-F(\rho _{0},\rho_{1})^{2}) & \tilde{F}\ge1
\end{array}
\]
This coincides precisely with the optimal solution derived in
\cite{SBH}. Thus, our lower bound is found to be optimal in this
case.

By recasting state filtering as an instance of UD of a pure state
and a mixed state, it has become apparent that the particular
ensemble of states from which the mixed state is formed (in this
case, $|\psi _{2}\>$ and $|\psi _{3}\>$ with priors $\eta_{2}$ and
$\eta_{3}$) is not significant. As such our analysis applies also
to the case wherein the second set of states contains an arbitrary
number of elements, in agreement with the solution found in
\cite{bergou}. Moreover, the mixed state need not even arise as a
result of ignorance of the pure state, but rather may arise as a
result of the system being entangled with another degree of
freedom. This is confirmed by the fact that the probability of
success depends only on $\rho_1$. The problem of UD between
multiple sets of pure states \cite{Zhang} is a straightforward
generalization of state filtering that can be recast as a problem
of UD of mixed states in a similar way.

\section{Conclusion}

We have shown that UD of a pair of mixed states is possible when
these have distinct supports. We have provided the maximum
probability of success for the case of mixed states with
orthogonal or 1-dimensional kernels. In the general case, we have
determined an upper bound based on the probability of UD for
purifications of the states, and a lower bound based on the
geometrical invariants between their kernels.  In addition to the
tasks discussed here, there are likely to be many others for which
UD of mixed states is useful, such as quantum random access codes,
quantum oblivious transfer and entanglement distillation, to name
a few.

%Furthermore, there are many optimization problems that are closely
%related to UD which should be generalized to the case of mixed
%states, such as determining the optimal discrimination for a given
%probability of error and determining the maximum probability of
%achieving a specified reduction in the fidelity between two
%states.  Note that whereas in the pure state case, after a
%successful discrimination, the discriminator can pass with
%certainty a test for the unknown state, in UD of \emph{mixed}
%states the discriminator is unable to pass, with certainty, some
%tests for the state -- namely tests for a purification of the
%state -- even when the discrimination was successful. We leave
%these interesting issues to future work.

\begin{acknowledgements} We thank F. Verstraete for useful
discussions, and S. Salmi for logistical support. TR is supported
by the NSA \& ARO under contract No. DAAG55-98-C-0040. RWS and PST
are supported in part by NSERC and the Sumner Foundation. RWS and PST wish
to unambiguously
distinguish themselves from TR, whose support is orthogonal to
their own.
\end{acknowledgements}

%\begin{references}

%\end{references}

\end{document}